\documentclass[a4paper, 10pt]{extarticle}
\usepackage{amsmath}
\usepackage{amssymb}
\usepackage{amsfonts}
\usepackage{graphicx}
\usepackage{subfigure}
\usepackage{authblk}
\usepackage[textwidth=16cm,textheight=20cm]{geometry}
\newtheorem{theorem}{Theorem}

\newenvironment{pf}[1][Proof]{\noindent\textbf{#1.} }{\ \rule{0.5em}{0.5em}}
\title{Phase transitions in filtration of Redlich-Kwong gases}
\author[1,2,\thanks{\textit{E-mail: }\texttt{valentin.lychagin@uit.no}}]{Valentin Lychagin}
\author[1,3,\thanks{\textit{E-mail: }\texttt{mihail\underline{ }roop@mail.ru}}]{Mikhail Roop}
\affil[1]{V.A. Trapeznikov Institute of Control Sciences, Russian Academy of Sciences, 65 Profsoyuznaya Str., 117997 Moscow, Russia}
\affil[2]{Department of Mathematics, The Arctic University of Norway, Postboks 6050, Langnes 9037, Tromso, Norway}
\affil[3]{Department of Physics, Lomonosov Moscow State University, Leninskie Gory, 119991 Moscow, Russia}
\date{}
\begin{document}
\maketitle

\abstract{
In this paper we study a 3-dimensional filtration of real gases described by Redlich-Kwong equations of state. Thermodynamical states are considered as Legendrian (Lagrangian) submanifolds in contact (symplectic) space. Connection between singularities of their projection on the space of intensive variables and phase transitions is shown. Explicit formulae for the Dirichlet boundary problem are given and the distribution of phases in space is shown.
}

\section{Introduction}
In this paper we study phase transitions in gas flows governed by nonlinear partial differential equations. The first results in this field were obtained in \cite{GLRT} applying methods developed in \cite{LY}. Here, we consider a 3-dimensional steady adiabatic filtration in porous media. The first results in this area were obtained in \cite{Lib,Mus}. They proposed to use the Darcy law instead of Navier-Stokes equations for this case. Filtration of ideal gases is considered in \cite{LychGSA}. Filtration processes in gases described by van der Waals and Peng-Robinson equations of state are studied in \cite{LRNon}.

Since the development of filtration processes takes a long time, we consider a steady filtration. This condition not only simplifies the mathematical model, but is also reasonable from the practical point of view.

The paper is organized as follows. First of all, we define thermodynamical states as Legendrian or Lagrangian manifolds in contact or symplectic space respectively (this description was first proposed in \cite{DLT}).  We mention (see \cite{LY} for details) that these manifolds are naturally equipped with differential quadratic form, which defines the applicable domains corresponding to different phases. The projections of submanifolds where this form changes its type are the curves separating the domains of applicability of our model, so-called \textit{spinodal curves}.  By means of the Massieu-Plank potential we provide the equations of \textit{coexistence curves}, i.e the curves where phase transition occurs. These methods are applied to one of the most popular gases in petroleum industry --- Redlich-Kwong gases. Particularly, we find the caloric equation for them and explain how to get coexistence curves for such gases.

The second part is devoted to filtration problem of real gases (see also \cite{LRNon}). We give explicit formulae for the Dirichlet boundary problems and discuss the case of point sources in detail. To illustrate these results we suppose that the medium is described by Redlich-Kwong model. For different numbers of point sources located on a plane we show how thermodynamical properties of such gases emerge along filtration process. Namely, we define the domains in space corresponding to different phases of Redlich-Kwong gases.

\section{Thermodynamical state}
In this section we briefly remind how ideas of contact and symplectic geometry \cite{KLR} can be applied in thermodynamics. More detailed description could be found in \cite{GLRT,LY,LRNon}.

First of all, thermodynamical states of gases are 2-dimensional Legendrian manifolds $\widehat{L}$ in contact space $(\mathbb{R}^{5},\theta)$ equipped with coordinates $(\sigma,p,T,v,e)$ standing for the specific entropy, the pressure, the temperature, the specific volume and the specific energy respectively, and structure form
\begin{equation*}
\theta=d\sigma-T^{-1}de-T^{-1}pdv,
\end{equation*}
i.e. $\theta|_{\widehat{L}}=0$. The projection $\pi\colon\mathbb{R}^{5}\to\mathbb{R}^{4}$, $\pi(\sigma,p,T,v,e)=(p,T,v,e)$ restricted on $\widehat{L}$ gives an immersed Lagrangian manifold $L$ in symplectic space $(\mathbb{R}^{4},\Omega)$ with structure form
\begin{equation*}\Omega=-d\theta=d(T^{-1})\wedge de+d(pT^{-1})\wedge dv,\end{equation*}
i.e. $\Omega|_{L}=0$.

Any Lagrangian surface $L\subset\mathbb{R}^{4}$ is defined by the two \textit{equations of state}
\begin{equation*}L=\left\{f(p,T,v,e)=0,\quad g(p,T,v,e)=0\right\},\end{equation*}
and the condition for $L$ to be Lagrangian may be expressed as follows:
\begin{equation*}[f,g]=0\text{ on }L,\end{equation*}
where $[f,g]$ is the Poisson bracket with respect to structure form $\Omega$:
\begin{equation*}[f,g]\hspace{1mm}\Omega\wedge\Omega=df\wedge dg\wedge\Omega.\end{equation*}

In thermodynamics the equations of state are usually of the form
\begin{equation*}f(p,T,v,e)=p-A(v,T),\quad g(p,T,v,e)=e-B(v,T).\end{equation*}
and the following theorem is valid \cite{LRNon}:
\begin{theorem}
The Lagrangian manifold $L$ for real gases is completely defined by \textit{Massieu-Plank potential} $\phi(v,T)$:
\begin{equation}\label{LegMani}p=RT\phi_{v},\quad e=RT^{2}\phi_{T}.\end{equation}
The specific entropy $\sigma$ and the specific Gibbs potential $\gamma$ have the following form:
\begin{equation}\label{Gibbs}\sigma=R(\phi+T\phi_{T}),\quad \gamma=RT(v\phi_{v}-\phi),\end{equation}
where $R$ is the universal gas constant.
\end{theorem}

The Lagrangian manifold $L$ is also equipped with the differential quadratic form $\kappa$ \cite{LY}:
\begin{equation*}\kappa=d(T^{-1})\cdot de+d(pT^{-1})\cdot dv.\end{equation*}
The set of points on $L$ where $\kappa$ is negative corresponds to applicable states. We call such points \textit{admissible}. In terms of Massieu-Plank potential we get
\begin{equation*}R^{-1}\kappa=-(\phi_{TT}+2T^{-1}\phi_{T})dT\cdot dT+\phi_{vv}dv\cdot dv,\end{equation*}
and the following theorem is valid:
\begin{theorem}
The domain of applicable states on the plane $(v,T)$ is given by inequalities
\begin{equation*}\phi_{vv}<0,\quad \phi_{TT}+2T^{-1}\phi_{T}>0.\end{equation*}
\end{theorem}
Note that due to state equations
\begin{equation*}e_{T}=RT(T\phi_{TT}+2\phi_{T}),\quad p_{v}=RT\phi_{vv}.\end{equation*}
Therefore, the domain of applicable states is given by inequalities
\begin{equation*}e_{T}>0,\quad p_{v}<0.\end{equation*}

Consider a submanifold $\Sigma\subset L$, which consists of the points, where differential form $dp\wedge dT$ has zeros, i.e.
\begin{equation*}\Sigma=\left\{\phi_{vv}=0\right\}.\end{equation*}
In this case the projection of the surface $L$ on the plane $(p,T)$ has singularities at $\Sigma$, applicable states are separated by $\Sigma$ from the set of points where $\phi_{vv}\ge0$ and a thermodynamical system does not exist. A jump from one admissible point $a_{1}=(p,T,v_{1},e_{1})\in L$ into another $a_{2}=(p,T,v_{2},e_{2})\in L$ governed by intensive variables $(p,T)$ and specific Gibbs free energy $\gamma$ conservation law $\gamma(a_{1})=\gamma(a_{2})$ is exactly what we call \textit{phase transition}. Points $a_{1}$ and $a_{2}$ we call \textit{phase equivalent}.

Phase equivalent points $(v_{1},T)$ and $(v_{2},T)$ could be found due to (\ref{LegMani}) and (\ref{Gibbs}). We get
\begin{equation}
\label{PhaseEqui}
\begin{split}
&\phi _{v}\left( v_{2},T\right)=\phi _{v}\left( v_{1},T\right),{}\\&
\phi \left( v_{2},T\right)- v_{2}\phi _{v}\left(
v_{2},T\right)= \phi \left( v_{1},T\right) -v_{1}\phi _{v}\left( v_{1},T\right).
\end{split}
\end{equation}
Equations (\ref{PhaseEqui}) allow to construct a coexistence curve in coordinates $(v_{1},v_{2})$ or, eliminating $v_{1}$ and substituting $v$ instead of $v_{2}$, in coordinates $(v,T)$.

Using equivalent equations
\begin{equation}
\label{PhaseEqui1}
\begin{split}
&\phi _{v}\left( v_{2},T\right)=\frac{p}{RT},\quad\phi _{v}\left(
v_{1},T\right) =\frac{p}{RT},{}\\&
\phi \left( v_{2},T\right)- v_{2}\phi _{v}\left(
v_{2},T\right)= \phi \left( v_{1},T\right) -v_{1}\phi _{v}\left( v_{1},T\right),
\end{split}
\end{equation}
it is possible to get a coexistence curve in $\mathbb{R}^{3}(p,v,T)$ and its projection on the plane $(p,T)$.

It is worth to say that both forms (\ref{PhaseEqui}) and (\ref{PhaseEqui1}) are essential for us, because they allow us to get coexistence curves in different coordinates and provide information about thermodynamical quantities on phase transition.

\section{Redlich-Kwong gases}
Redlich-Kwong equation of state was proposed by O. Redlich and J.S.N. Kwong in \cite{RedKw}. This is a two-parameter equation of state and it appeared to give adequate results in description of non-polar hydrocarbons. That is why this equation is of wide use in petroleum industry.

The first state equation for Redlich-Kwong gases has the following form:
\begin{equation*}f(p,T,v,e)=p-\frac{RT}{v-b}+\frac{a}{\sqrt{T}v(v+b)},\end{equation*}
where $a$ and $b$ are constants depending on the gas.

To find the second equation $g(p,T,v,e)=0$ we assume that $g(p,T,v,e)=e-\beta(v,T)$ and take the Poisson bracket $[f,g]$ with respect to structure form $\Omega$. The restriction of this bracket on $L$ has to be equal to zero, and we get an equation for function $\beta(v,T)$:
\begin{equation*}
3a-2v\sqrt{T}(v+b)\beta_{v}=0.
\end{equation*}
It's solution is
\begin{equation*}
\beta(v,T)=F(T)+\frac{3a}{2b\sqrt{T}}\ln\left(\frac{v}{v+b}\right).
\end{equation*}
Since in case of $a=0$, $b=0$ we get ideal gas, we have to define $F(T)=nRT/2$, where $n$ stands for the degree of freedom.

The specific entropy for Redlich-Kwong gases is
\begin{equation*}
\sigma(v,T)=\frac{Rn}{2}\ln T+R\ln(v-b)+\frac{a}{2bT^{3/2}}\ln\left(\frac{v}{v+b}\right).
\end{equation*}
Thus, the Legendrian manifold $\widehat{L}$ for Redlich-Kwong gases is determined completely.

Let's introduce the following scale contact transformation:
\begin{equation*}p\mapsto\left(\frac{Ra^{2}}{b^{5}}\right)^{1/3}p,\quad T\mapsto\left(\frac{a}{Rb}\right)^{2/3}T,
\quad v\mapsto bv,\quad e\mapsto\left(\frac{Ra^{2}}{b^{2}}\right)^{1/3}e,\quad\sigma\mapsto R\sigma.\end{equation*}

Then the state equations for Redlich-Kwong gases take the following form in new coordinates, which we continue denoting by $p,T,v,e$:
\begin{equation*}p=\frac{T}{v-1}-\frac{1}{\sqrt{T}v(v+1)},\quad e=\frac{nT}{2}+\frac{3}{2\sqrt{T}}\ln\left(\frac{v}{v+1}\right).\end{equation*}
Note that because of positivity of $p$ and $T$ only $v>1$ have a physical meaning.

The specific entropy and the Massieu-Plank potential are of the form:
\begin{eqnarray}
\label{entrRed}
\sigma(v,T)&=&\frac{n}{2}\ln T+\ln(v-1)+\frac{1}{2T^{3/2}}\ln\left(\frac{v}{v+1}\right),\\
\phi(v,T)&=&\frac{n}{2}\ln T+\ln(v-1)-\frac{1}{T^{3/2}}\ln\left(\frac{v}{v+1}\right),
\end{eqnarray}
and the differential quadratic form $\kappa$ has the following structure on $L$:
\begin{eqnarray*}\kappa R^{-1}=&-&\left(\frac{n}{2T^{2}}+\frac{3}{4T^{7/2}}\ln(1+v^{-1})\right)dT\cdot dT\\&-&\frac{v^{2}(v+1)^{2}T^{3/2}-2v^{3}+3v^{2}-1}{T^{3/2}(v+1)^{2}v^{2}(v-1)^{2}}dv\cdot dv.\end{eqnarray*}

\begin{figure}[h!]
\centering
\includegraphics[scale=.3]{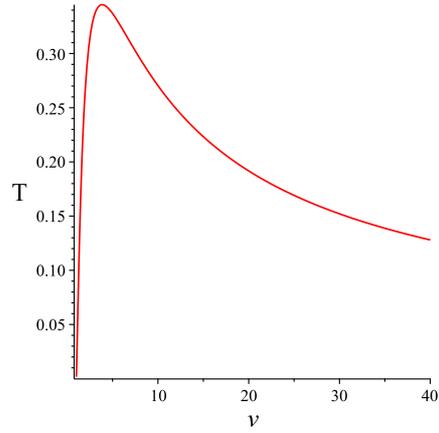}
\caption{Spinodal for Redlich-Kwong gases. The domain above the curve corresponds to applicable states.}
\label{applicRK}       
\end{figure}

Note, that comparing with van der Waals gases and Peng-Robinson gases (see\cite{GLRT,LRNon}) component $\kappa_{11}$ depends on the specific volume. But since $v>1$, $\kappa_{11}<0$ at any point on the Lagrangian surface, which means that the projection of $L$ on the space of extensive variables $(e,v)$ has no singularities \cite{LY}. But because of sign changing of $\kappa_{22}$ the projection of $L$ on the plane $(p,T)$ has singularities. The corresponding spinodal is shown in figure~\ref{applicRK}.

\subsection{Coexistence curves}
\begin{figure}[h!]
\centering
\includegraphics[scale=.3]{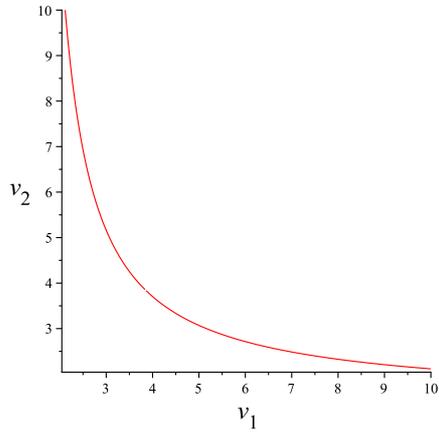}
\caption{Coexistence curve for Redlich-Kwong gases.}
\label{coexV}       
\end{figure}

As we have shown, the coexistence curve for Redlich-Kwong gases can be obtained by means of Massieu-Plank potential. Eliminating $T$ from equations~(\ref{PhaseEqui}) we get a coexistence curve in $\mathbb{R}^{2}(v_{1},v_{2})$ (see figure~\ref{coexV}), which shows the specific volumes of phase transition.

\begin{figure}[ht!]
\centering \subfigure[]{
\includegraphics[width=0.4\linewidth]{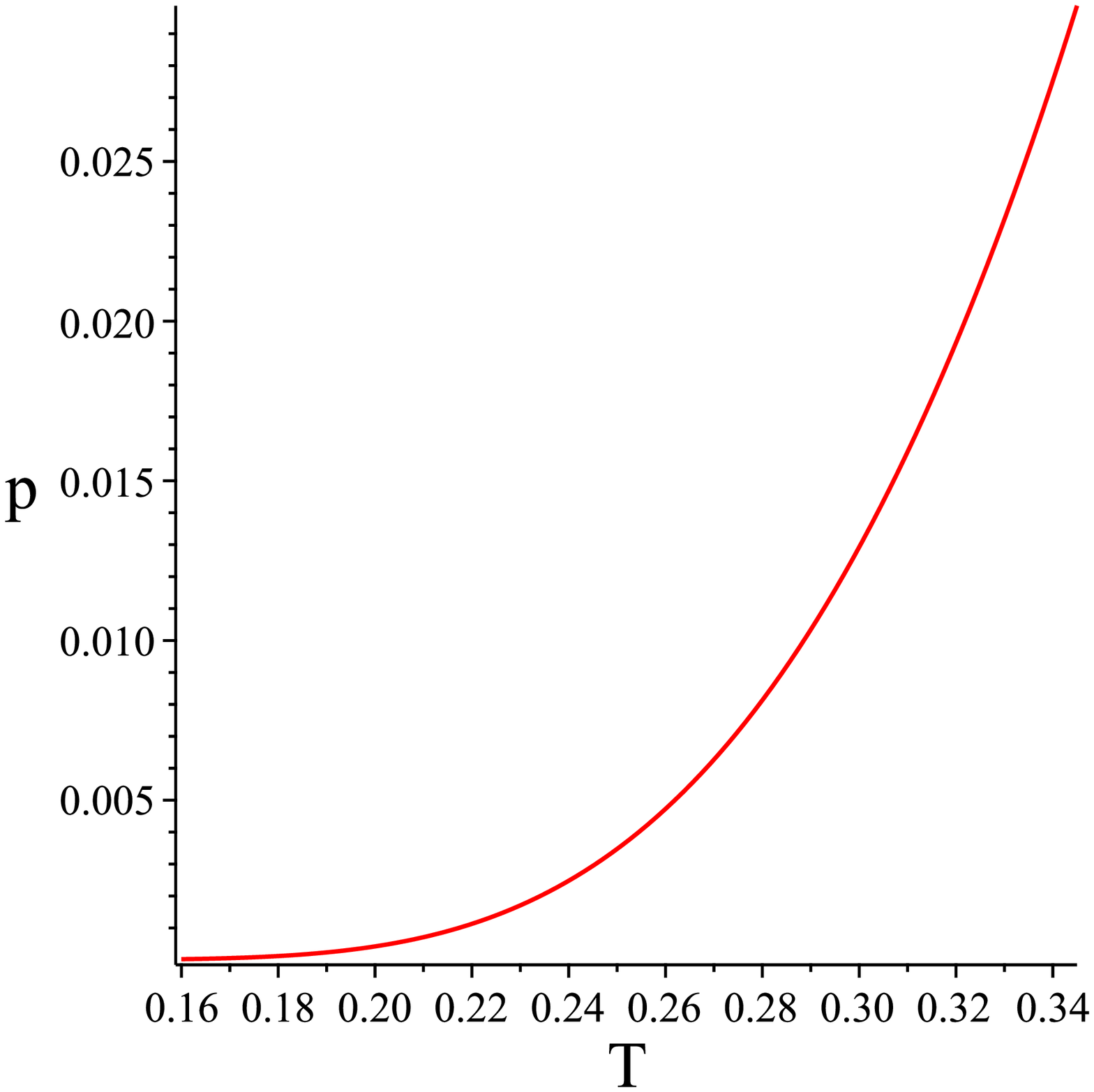} \label{curvepT} }
\hspace{4ex}
\subfigure[]{ \includegraphics[width=0.4\linewidth]{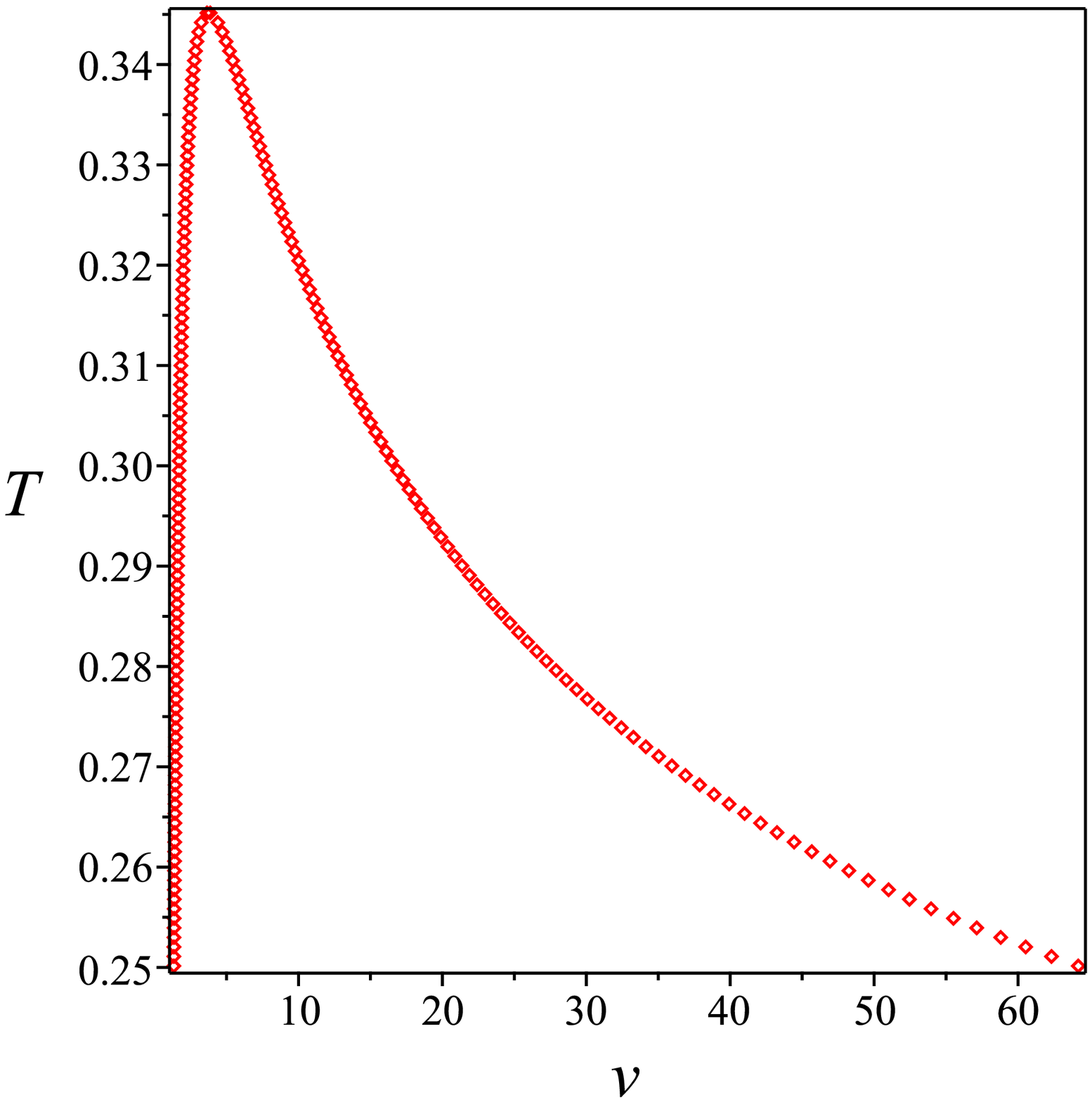} \label{curvevT} }
\caption{Coexistence curves for Redlich-Kwong gases: \subref{curvepT} in $\mathbb{R}^{2}(p,T)$, liquid phase is on the left of the curve, gas phase is on the right; \subref{curvevT} in $\mathbb{R}^{2}(v,T)$, inside the curve --- intermediate state (condensation process).} \label{CoexCurves}
\end{figure}

By means of~(\ref{PhaseEqui1}) and the equations of state, we can get coexistence curves in different coordinates, eliminating corresponding thermodynamical variables. In coordinates $(p,T)$  and $(T,v)$ they are presented in figures \ref{curvepT} and \ref{curvevT} respectively. However, it is quite complicated to obtain explicit formulae, for example, for $p$ as function of $T$. For this reason we construct such curves numerically.

\begin{figure}[h!]
\centering
\includegraphics[scale=.3]{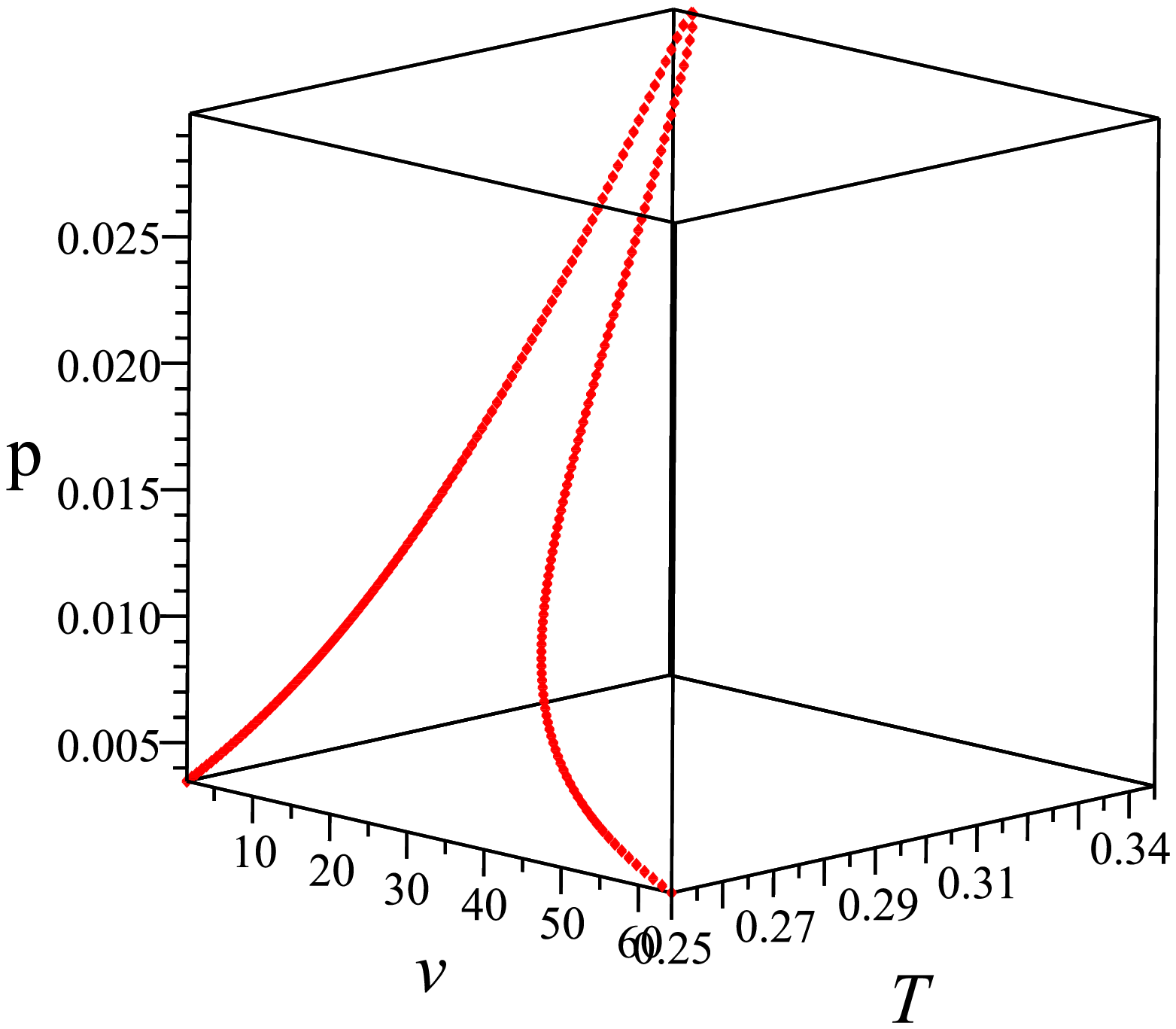}\\
\caption{Coexistence curve for Redlich-Kwong gases in $\mathbb{R}^{3}(p,v,T)$ .}
\label{curvepvT}
\end{figure}

These curves can be lifted into the space $\mathbb{R}^{3}(p,v,T)$, which is shown in figure~\ref{curvepvT}.

\section{Steady adiabatic filtration of real gases}
\subsection{Equations for isotropic filtration}
Steady filtration of gases in 3-dimensional homogeneous isotropic porous media is described by the following system of differential equations \cite{Lib,Mus,Sch}:
\begin{equation}
\label{BasicEqs}
\left\{
\begin{aligned}
&\mathrm{div}(\rho\mathbf{u})=0,\\
&\mathbf{u}=-\frac{k}{\mu}\nabla p,\\
&(\mathbf{u},\nabla\sigma)=0,
\end{aligned}
\right.
\end{equation}
where $\mathbf{u}(x)$ is the vector field of filtration rate, $p(x)$ is the pressure, $\rho(x)=v^{-1}(x)$ is the density, $\sigma(x)$ is the specific entropy, $x=(x_{1},x_{2},x_{3})\in D\subset\mathbb{R}^{3}$, $k$ is the permeability coefficient depending on medium, $\mu$ is the dynamic viscosity. The first equation in (\ref{BasicEqs}) corresponds to the mass conservation law, the second one is the Darcy law and the third equation is the condition of the specific entropy constancy along the trajectories of $\mathbf{u}$.

Condition $(\mathbf{u},\nabla\sigma)=0$ in case of sources leads to the constancy of the specific entropy in neighborhoods of the sources. Consider a domain $D\subset\mathbb{R}^{3}$ with sources $\left\{a_{i}\right\}$. The domain $D$ can be presented as a union of domains $D_{k}$, such that sources in $D_{k}$ have the same entropy. Filtrations in $D_{k}$ are independent. For this reason we restrict our consideration on the case of domains with constant specific entropy $\sigma=\sigma_{0}$.

Assume that fixed level of the specific entropy is given. Then all the thermodynamical values can be expressed as functions of the specific volume $v$. Indeed, using equations of state we get Massieu-Plank potential $\phi$ as function of temperature $T$ and specific volume $v$ and due to (\ref{Gibbs}) we have the following equation:
\begin{equation}\label{eqfortemp}\phi +T\phi _{T}=\sigma _{0},\end{equation}
which defines $T(v)$. There exists a solution $T(v)$ of (\ref{eqfortemp}), because the derivative of left hand side with respect to $T$ is positive in applicable domain. Using equation of state we get $p(v)$. Define a function $Q(v,\sigma_{0})$ in the following way:
\begin{equation*}Q(v,\sigma_{0})=-\int\frac{k}{v\mu}p^{\prime}(v)dv.\end{equation*}
\begin{theorem}Basic equations (\ref{BasicEqs}) of adiabatic filtration are equivalent to equation
\begin{equation}\label{eqforQ}\Delta(Q(v,\sigma_{0}))=0,\end{equation}
where $\Delta$ is the Laplace operator.
\end{theorem}
\begin{pf}
Indeed,
\begin{equation*}\begin{split}
0&=\mathrm{div}(v^{-1}\mathbf{u})=-\mathrm{div}\left(\frac{k}{\mu v}\sum\limits_{i}p_{x_{i}}\frac{\partial}{\partial x_{i}}\right)=\\{}&=\mathrm{div}\left(\sum\limits_{i}-k\frac{p^{\prime}(v)}{\mu v}\frac{\partial v}{\partial x_{i}}\frac{\partial}{\partial x_{i}}\right)=\mathrm{div}\left(\sum\limits_{i}Q_{x_{i}}\frac{\partial}{\partial x_{i}}\right)=\Delta Q.
\end{split}\end{equation*}
\end{pf}

Thus, due to above theorem we have solutions for $v$ and $T$ as functions in $D$. Substituting them to equations defining coexistence curves, we get domains in $D$ where phase transitions occur.
\subsection{Dirichlet problem}
Here, we describe the method of finding solutions in case of point sources \cite{LRNon}. Due to (\ref{eqforQ}) we should take a harmonic in a domain $D$ function $u(x)$ and the relation
\begin{equation*}u(x)=Q(v,\sigma_{0})\end{equation*}
defines in general multivalued function $v(x)$, satisfying equations (\ref{BasicEqs}).

Consider an open and connected domain $D\subset\mathbb{R}^{3}$ with a smooth boundary $\partial D$ and let $A=\left\{a_{i}, i=\overline{1,N}\right\}\subset{D}$ be a set of points where the sources with given intensities $J_{i}$ are located. We are looking for the solution $v$ of Dirichlet problem of basic equations in domain $D\setminus A$:
\begin{equation*}\Delta(Q(v))=0,\quad v|_{\partial D}=v_{0}.\end{equation*}
 In this case we should take a harmonic in $D\setminus A$ function $u(x)$ of the form
 \begin{equation*}u=\sum\limits_{i=1}^{N}\frac{J_{i}}{4\pi|x-a_{i}|}+u_{0},\end{equation*}
 where $u_{0}$ is a harmonic in $D$ function with boundary conditions:
 \begin{equation*}u_{0}|_{\partial D}=Q(v_{0})-\left.\left(\sum\limits_{i=1}^{N}\frac{J_{i}}{4\pi|x-a_{i}|}\right)\right|_{\partial D}.\end{equation*}
Taking $Q^{-1}(u)$ we get in general multivalued solution $v(x)$ of the Dirichlet boundary problem.

\subsection{Redlich-Kwong gases filtration}
Here, we consider adiabatic filtration of Redlich-Kwong gases in case of a number of sources. The coefficients $k$ and $\mu$ are assumed to be constants.

From above saying follows that to construct a single-valued solution we need the invertibility conditions of function $Q(v,\sigma_{0})$, or, in other words, find the conditions when function $Q(v,\sigma_{0})$ is monotonic. We consider $\sigma_{0}$ as a parameter. This problem can be reformulated as follows. We need to find a specific entropy level $\sigma_{0}$, such that $Q_{v}(v,\sigma_{0})\ne 0$ if $v>1$. Since the conditions $Q_{v}(v,\sigma_{0})=0$ and $p_{v}(v,\sigma_{0})=0$ are equivalent, we need an explicit expression $p(v,\sigma_{0})$.

Note that the following relation holds due to~(\ref{entrRed}):
\begin{equation}
\label{entrlev}
\sigma_{0}=\frac{n}{2}\ln T+\ln(v-1)+\frac{1}{2T^{3/2}}\ln\left(\frac{v}{v+1}\right).
\end{equation}
We cannot write an explicit formula $T=T(v,\sigma_{0})$ and $p=p(v,\sigma_{0})$ as well. However, we can estimate asymptotic behavior for $T(v,\sigma_{0})$, $p(v,\sigma_{0})$ and $Q(v,\sigma_{0})$ for $n=3$.
\begin{theorem}
If $v\to +\infty$ then asymptotics for $T(v,\sigma_{0})$, $p(v,\sigma_{0})$ and $Q(v,\sigma_{0})$ have the following form:
\begin{equation*}\begin{split}&T(v,\sigma_{0})=\frac{1}{(B^{*}v)^{2/3}}+O\left(\frac{1}{v^{5/3}}\right),\quad p(v,\sigma_{0})=\frac{c}{v^{5/3}}+O\left(\frac{1}{v^{8/3}}\right),{}\\&
Q(v,\sigma_{0})=-\frac{5kc}{8\mu v^{8/3}}+O\left(\frac{1}{v^{11/3}}\right),
\end{split}\end{equation*}
where $B^{*}$ is the root of the equation
\begin{equation*}
-\sigma_{0}=B/2+\ln B,
\end{equation*}
and
\begin{equation*}
c=\left(B^{*}\right)^{-2/3}-\left(B^{*}\right)^{1/3}.
\end{equation*}
\end{theorem}

\begin{theorem}
If $v\to 1$ then asymptotics for $T(v,\sigma_{0})$, $p(v,\sigma_{0})$ and $Q(v,\sigma_{0})$ have the following form:
\begin{equation*}\begin{split}&T(v,\sigma_{0})=\frac{B^{2/3}}{(v-1)^{2/3}}+O\left((v-1)^{1/3}\right),\quad p(v,\sigma_{0})=\frac{B^{2/3}}{(v-1)^{5/3}}+O\left(\frac{1}{(v-1)^{2/3}}\right),{}\\&
Q(v,\sigma_{0})=-\frac{kB^{2/3}}{\mu (v-1)^{5/3}}+O\left(\frac{1}{(v-1)^{2/3}}\right),
\end{split}\end{equation*}
where $B=e^{\sigma_{0}}$.
\end{theorem}

Note that due to equations of state $p=p(T,v)$ and
\begin{equation}
\label{dpdv}
\frac{dp}{dv}=\frac{\partial p}{\partial v}+\frac{\partial p}{\partial T}\frac{dT}{dv}.
\end{equation}
We can find $T^{\prime}(v)$ by means of~(\ref{entrlev}) and substitute it in~(\ref{dpdv}). Since $p^{\prime}(v)=0$, we get an expression
\begin{equation*}G(T,v)=0,\end{equation*}
which defines the relation between temperature $T$ and specific volume $v$ when $Q_{v}=0$. This expression is too large to write it in this paper, but it can be resolved with respect to $T$ explicitly. The root $T(v)$ we substitute in~(\ref{entrlev}) and get an equation
\begin{equation*}
\sigma_{0}=H(v).
\end{equation*}
If for any $v>1$ the above equation has a solution, function $Q(v)$ is irreversible and we have a number of possibilities in filtration development. Otherwise, thermodynamical properties are uniquely determined.

\begin{figure}[h!]
\centering
\includegraphics[scale=.25]{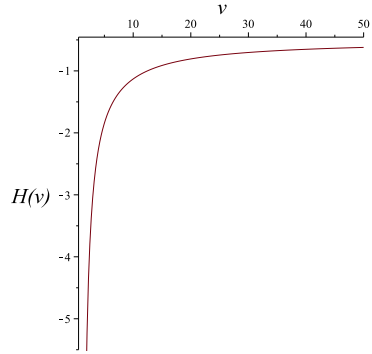}
\caption{Graph of function $H(v)$.}
\label{funcH}       
\end{figure}

For $n=3$ the graph of function $H(v)$ is presented in figure~\ref{funcH}. As we can see, it has a limit when $v\to\infty$, which can be computed numerically and it equals $\sigma^{*}=-0.5$. So, if $\sigma_{0}>\sigma^{*}$, function $Q(v)$ is invertible.

\begin{figure}[ht!]
\centering \subfigure[]{
\includegraphics[width=0.4\linewidth]{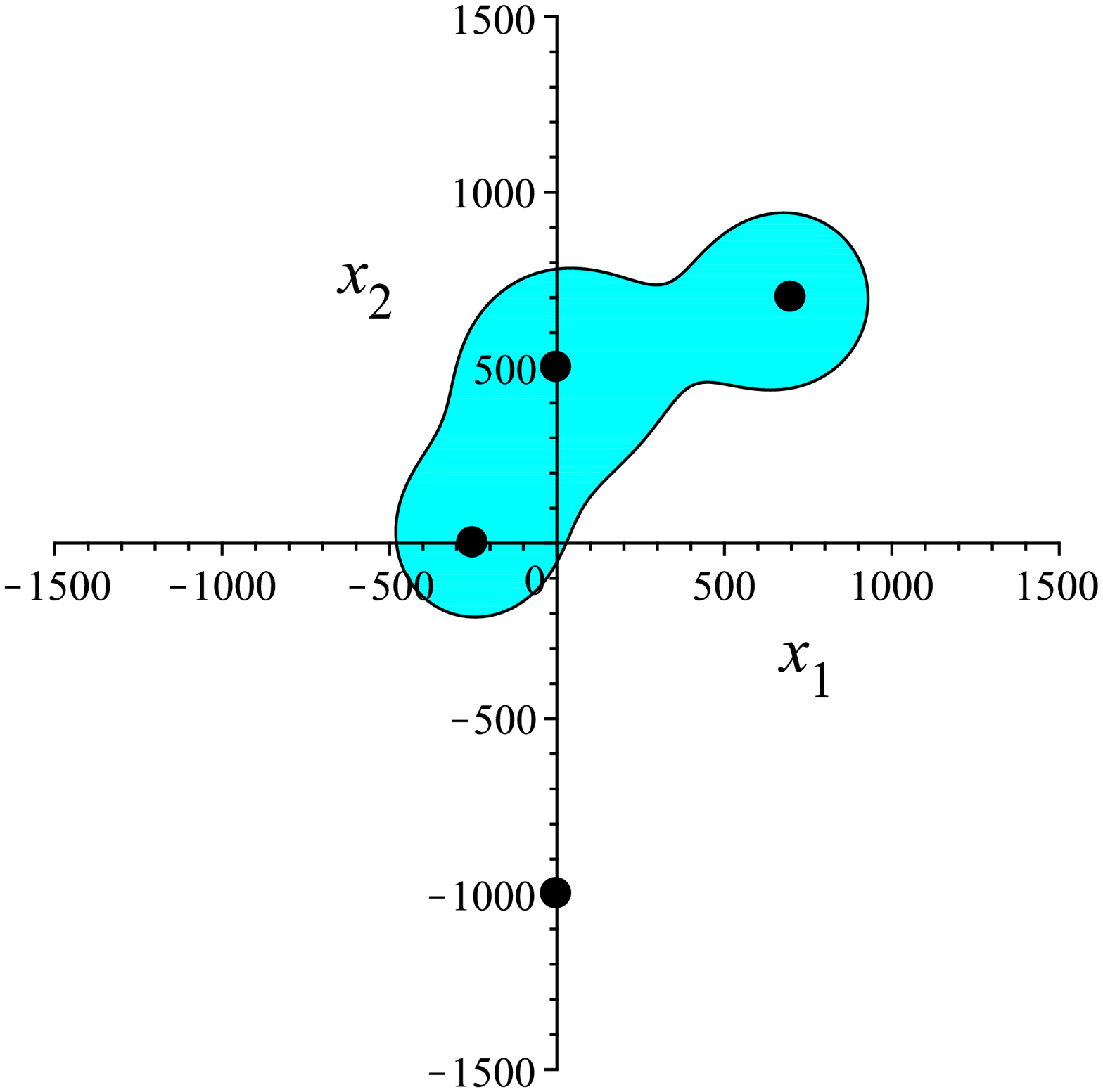} \label{sources4} }
\hspace{4ex}
\subfigure[]{ \includegraphics[width=0.4\linewidth]{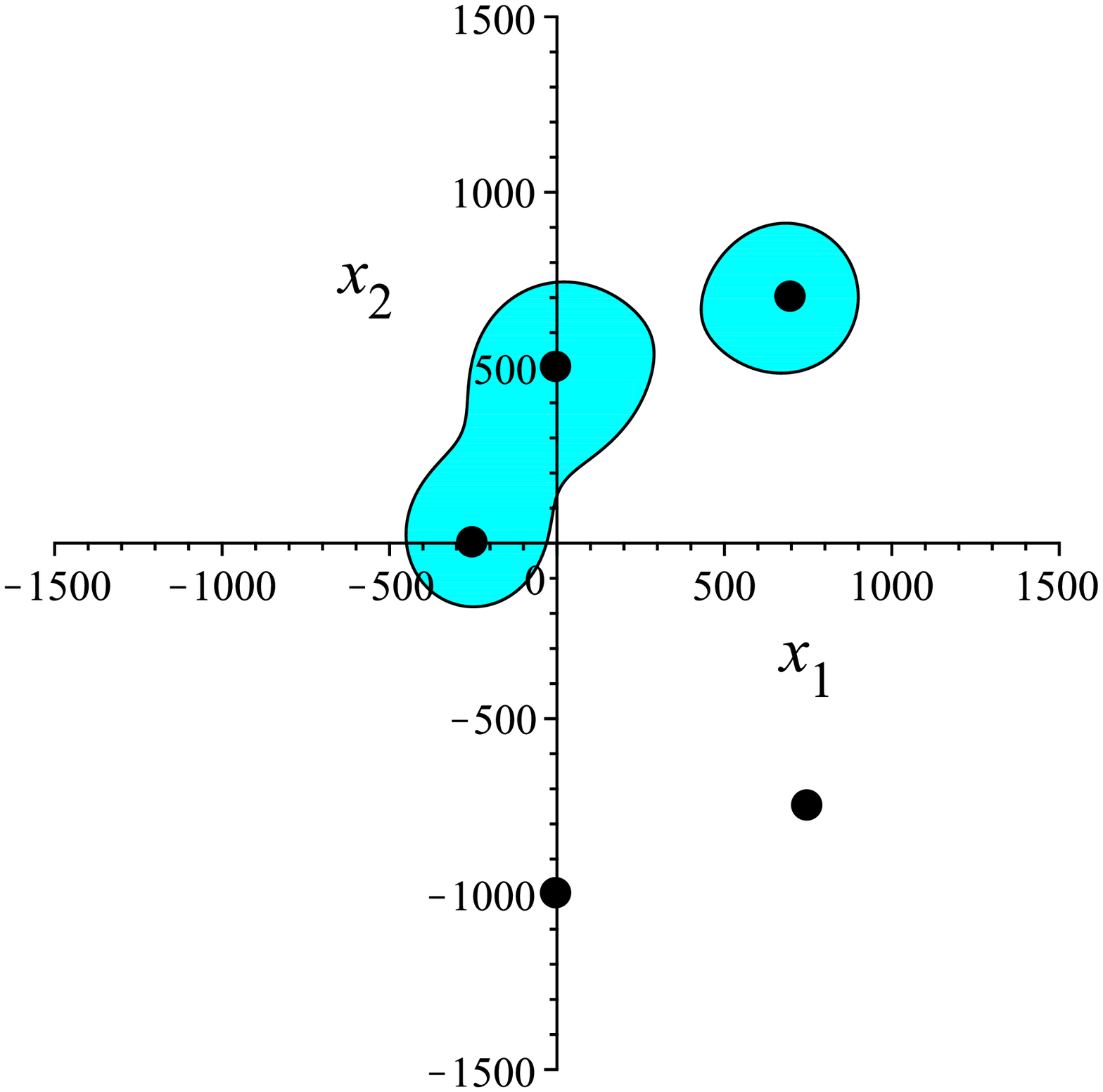} \label{sources5} }
\caption{The distribution of phases in space for Redlich-Kwong gases: \subref{sources4} in case of $N=4$ sources; \subref{sources5} in case of $N=5$ sources. Black points are the sources, white domain corresponds to gas phase, coloured domain is the condensation process.} \label{phaseSpace}
\end{figure}

Suppose that we have a number of sources with different intensities located on the plane $x_{3}=0$. The distribution of phases in space for this case is shown in figure \ref{phaseSpace}.

\section*{Acknowledgements}
This work was supported by the Russian Foundation for Basic Research (project 18-29-10013).

\end{document}